**Chechelnitsky A.M.**

# PHANTOM OF HIGGS BOSON VERSUS HIERARCHY OF STATIONARY STATES OF SUPERHIGH ENERGIES

**Chechelnitsky A.M.**

# Phantom of Higgs Boson Versus Hierarchy of Stationary States of Superhigh Energies

Laboratory of Theoretical Physics,
Joint Institute for Nuclear Research,
141980 Dubna, Moscow Region, Russia
E'mail: ach@thsun1.jinr.ru

**Chechelnitsky Albert Michailovich –**

is an astrophysicist, cosmologist,

expert in space research, theoretical physics, theory of dynamic systems,

automatic control, optimization of large systems,

econometrics, constructive sociology, anthropology;

COSPAR Associate: Member of International organization – Committee on Space Research (COSPAR) – Member of B, D, E Scientific Commissions.

(COSPAR – most competent international organization, connected with fundamental interdisciplinary investigations of Space).

Author of the (Mega) Wave Universe Concept.

**Chechelnitsky A.M.**

**Phantom of Higgs Boson Versus Hierarchy of Stationary States of Superhigh Energies.**

## ABSTRACT

As is known, the Standard Model mainly ideologically and qualitatively focuss the experimenters in their search of new mass states (of EP- elementary particles). The exact quantitative prognosis of their properties, especially of masses, lays outside opportunities of the usual theory.

Model of Stationary states of EP within the framework of the Wave Universe Concept [Chechelnitsky, 1980-2001] points on existence of Hierarchy of physically distinguished - stationary (elite, dominant) states described by the mass formulas, in particular, in a range 10÷210 Gev/$c^2$:

The states close to…, 101.5; 107.3; 112.76÷113; 139.5÷143; 147.6; 202 Gev/$c^2$ should be observed.

Apparently, the experiment already confirms this prognosis in a range up to 100 Gev/$c^2$. You see preferable states, observable already now in experiment, it - not rejected by the usual theory as the candidates in constituents of Standard model (for example, not holding Higgs bosons), but quite real displays of stationary (first of all, - dominant) mass states.

Last data of L3 (CERN) Collaboration really specify displays of new mass states and close to 103.7; 108.9; 114.5 Gev/$c^2$.

March 2001

## RANGE 12-210 Gev/c$^2$: "GREAT DESERT"?

On a boundary of centuries formed in physics of high energies situation is characterized by some remarkable circumstances:

# According to prevailing representations of Standard model the extreme efforts of physical community are still concentrated on search and judgement of postulated basic constituents of the theory – of bosons, realizing weak interaction ($Z^0$ - neutral currents, $W^{\pm}$ - charged currents) and of Higgs bosons $H^0$, responsible for observable in experiments manifestation of real masses.

# It is considered, that observable mass states in area of masses M=80.3 Gev/c$^2$, M=91.2 Gev/c$^2$ is just those carriers (quantums) of interaction, which are responsible for weak interaction.

# Nowadays at the centre of attention there is a search of Higgs boson $H^0$ [Felcini, 2000; Tully, 2000; Acciarri et al., 2000].

# As the theory essentially is not capable precisely to point, in particular, major parameter – mass of $H^0$ boson (as however – and mass of top-quark), the search is conducted by the tested way - "*At random*".

# This "method" already is compelled was applied at total "combing" of a range M=80.3 Gev/c$^2$, resulted to detection of mass configurations in area M=80.3 Gev/c$^2$, M=91.2 Gev/c$^2$.

You see here again there were no precise and exact indications of the theory on localization of required states on a scale of masses.

# The rather specific characteristics of postulated Higgs boson $H^0$ result to hard selection of the potential candidates.

Many *really observable* mass states are rejected as unsufficiently valid the candidates in $H^0$ bosons.

By virtue of a similar sort of the factors, traditions and preferences in representation of physical community there is a following picture of HEP on a boundary of centuries.

# From 10 up to 210 Gev/c$^2$ and, probably, further “Great Desert " reaches, where, as it is considered, there are no mass states, deserving attention.

# Above it the peaks $W^{\pm}$, $Z^0$ of bosons tower as area of Everest only.

# Till now fruitless search of Higgs boson is conducted (for the present).

# But still adepts of Standard model are complete of optimism. As is known, - the theorists frequently are mistaken, but never doubt.

**Crisis of Belief.**

We suppose, that such picture, dictating by preferences of the prevailing theory, in many respects, will *disorient* not only theorists, but, main, - the experimenters conducting intense, extremely difficult search in a rich fog of unverified opinions, were guided only by assurances of authorities.

The situation too obviously reminds fantastic "Go there - I do not know where". But main, - the dominated dogmas of habitual representations *extremely narrow prospects of experimental search.*

There is a hunt only at *widely known* Phantoms of the prevailing theory.

**Other Horizons.**

Other prospects are offered by system of representations connected with Wave Universe Concept (WU Concept) [Chechelnitsky (1978) 1980-2000].

As against Standard model not capable to point exactly, for example, localization of new states on a scale of mass, the offers of WU Concept are *rather critical*.

The offered mass spectrum of new stationary states [Chechelnitsky, 2000] is *quite certain* and is *unequivocal.* It can be veryfied- confirmed or denied by an obvious way by experiment.

**Panorama Represented by an $G^{[2]}$ Shell.**

The picture offered by WU Concept for forward edge of HEP is rather certain [see also Chechelnitsky, 2000].

# There is a whole set of *physically distinguished - elite* (among them - strongest - *dominant*) states in area of masses

$$M = 10 - 210 \text{ Gev/c}^2.$$

# This cluster of stationary states we shall present by an $G^{[2]}$ Shell. *The mass spectrum* of dominant (elite) states is

described by the *Mass formulas* for stationary states [Chechelnitsky, 2000] (see. the Tables 1,2,3).

# In a range up to M ~100 Gev/$c^2$ should be observed the *dominant* states in area

12.6, 17.8÷19.5, 25.2, 35.7÷36.5, 50.4, 71.4÷71.8, 76.8 Gev/$c^2$.

# Separate theme - *true* physical sense and nature of detected in experiment states with masses close to M = 80.84 Gev/$c^2$ and to M = 91.2 Gev/$c^2$. Its - states laying close to elite values of the Main quantum number N = 16.5 and 17.5.

# On periphery of an $G^{[2]}$ Shell in area of masses 100 - 200 Gev/$c^2$ extend the *Transitive zone* (it - dynamic analogue of Transitive zones of asteroids and comets in Shells $G^{[1]}$ and $G^{[2]}$ of Solar system [Chechelnitsky,1986,1992,1999]). It is necessary to expect, that detected in this range the mass states will be, generally speaking, less steady, than states in another (with smaller masses) half of $G^{[2]}$ Shell.

**Perspective Search.**

# New Renessance – manifestation of new mass states (following behind a Transitive zone of fading, less steady states in an $G^{[2]}$ Shell) it is necessary to expect at detecting of physically distinguished - (elite) dominant states in area of the *following* $G^{[3]}$ *Shell.* It is a range of masses

$$M = 170 - 2700 \text{ Gev}/c^2.$$

# Mass spectrum of dominant states laying in the subsequent range

$$M = 2.28 \div 36.5 \text{ Tev}/c^2$$

is represented by an $G^{[4]}$ Shell.

All this - perspective field of researches of HEP of new century.

The spectrum of potentially arising mass configurations, which will be met by the experimenters in forthcoming search, is described by the mass formulas of WU Concept for stationary states (see. the Tables 1,2,3).

**Reference Points for Experiment.**

As is known, the basic lesson of a History (science, including) is, that nobody takes from it of the special lessons. Nevertheless, we shall try to comprehend the future.

# The experimenters substantially will facilitate to themselves life, and, main, will achieve decisive results, if, whenever possible, get rid from tyrannical influence of habitual dogmas of the settled theory (Standard model).

You see, - on the one hand, it does not give the exact instructions, where (in what place on a scale of masses) to search, for example, Higgs boson (top-quark, etc).

On the other hand, it approves, that it is *not enough* of the required candidates (on a role of Higgs boson) ab definito.

# Tactics, used by the experimenters, of *severe extreme selection* from here follows. And consequently, it is ((probable) *quite* steady mass states are exposed to rather rigid selection, are denied already during experiments and its are wrongful eliminated from a field of consideration of the experimenters and independent theorists.

# As against the usual representations, WU Concept approves, that in a range $M=10\div210$ Gev/$c^2$ (and higher) *there is a rather advanced Hierarchy of the physically distinguished states*, on a variety, probably, not yielding to observable in experiment Hierarchy of states in a range up to $M=10$ Gev/$c^2$ (nowadays – to basic contents of Particle Data Group).

# At presence of such polar, alternative representations it is best to the experimenters *to not trust finally anybody*, but to give steadfast attention to *each* of mass states, opening in experiment, - *without preliminary theoretical selection and imposed assumptions*.

# Received the advanced experimental spectrum of the physically distinguished states (*all - bar none*) will ensure, except for other, also *objective verification* of competing theoretical models.

**The Future History of HEP.**

Analyzing the latent tendencies, social aspects and human, psychological motives accompanying to development of exact sciences - cosmology, physics, including, - physics of high energy (HEP), it is possible to imagine and picture of the future development of HEP. The enormous, extreme intellectual and material efforts persistently *require the subsequent justification*. The social order is those.

The physical community can not permit itself to admit that the caravan of highly experimentally equipped science long time moved not in right direction or has lost the way because the theory in next time gave the incorrect instructions.

The expectations and searches of the justification are so great, that, is possible, the Higgs boson will be, by and large, "open" in foreseeable time. Suitable, the experimentally found out mass state can be soon successor of "empty throne " and will be coronated (and interpreted) by the prevailing theory in the main generator (the Higgs mechanism, "moderator" or "Emperor") of masses.

So, actually, has taken place and with experimentally observable mass states in area M=80.8 Gev/$c^2$ and M=91.2 Gev/$c^2$. Its were *announced* (are interpreted) as so long and iintense expected by the theory (quantums) of weak interaction - $W^{\pm}$ and $Z^0$ bosons - with all accompanying attributes, activity of scientific and social mass-media and "with distribution of prizes and elephants ".

In contrast with it there are serious bases to believe, that

# *Do not exist* the postulated by the Standard theory the special Higgs mechanism and appropriate (set) of Higgs bosons.

Such thought up the ad hoc concept and connected with it constituents ($H^0$, etc) are not by the necessary functional basis of the effective, serviceable theory. Especially as such concepts do not follow from *dynamic* and physical base principles.

# (As) *there are no* carriers (quantums) of *weak* interaction laying in area of *high* masses (and energies) M=80.8 Gev/$c^2$ and M=91.2 Gev/$c^2$. Such the conceptual inversion is hardly viable and is hardly realized in Hierarchy of masses and interactions in the Universe.

# The states with masses M=80.8 Gev/$c^2$ and M=91.2 Gev/$c^2$ are *quite independent* and, generally speaking,

*ordinaries* states - same as many other of compendium of the data of Particle Data Group [RPP] [see the Tables 1,2,3].

**What is Farther? New Horizons.**

Peering in the Future, it is necessary to hope, that the boundary of centuries will appear also time *of deep, critical* doubts and choice of *new ways*, with which the physics of new time will follow.

Is possible, as a result of the severe analysis that the physics of high energies long time went in a fog, following for phantoms externally attractive, for fantastically beautiful constructions of the usual theory.

But when the fog of biases eventually will dissipate, we shall see completely other bright picture of HEP, sated by *set* of mass states (resonances) demonstrating an *advanced spectrum of Hierarchy.*

Brightest of them will correspond to physically distinguished – *dominant* states of the *Fundamental spectrum (of masses) of stationary states*.

This spectrum arises not in result each time again thought out ad hoc mechanisms and theories. Were based on fundamental principles, Nature, the Wave Universe with use only of *simple and universal* receptions in *recurrent* regime builds *all observable Hierarchy* of the physically distinguished states of micro - and megaworld.

Table 1

## MASS SPECTRUM: STATIONARY STATES - $G^{[2]}$ SHELL

| **THEORY** | | | | | | **EXPERIMENT** |
|---|---|---|---|---|---|---|
| **Micro – Mega (MM) Analogy** | | | **General Dichotomy** | | | **Experiment** [RPP,1994,p.1367;RPP,2000] |
| **States** | **Quantum Number** **N** | **Mass** **M=** **$M_*(N^2/2\pi)$** **$M_*$=1.8675** | **States** **ν** | **Quantum Number** **$N=N_\nu= N_{\nu=0}2^{\nu/2}$,** **$N_{\nu=0}$=6.5037** | **Mass** **M=** **$M_*(N^2/2\pi)$** **$M_*$=1.8675** | **Mass** **M** |
| | | **[Gev/c$^2$]** | | | **[Gev/c$^2$]** | **[Gev/c$^2$]** |
| **$TR_*$** | 2.5066 | 1.8675 | | 2.5066 | 1.8675 | |
| | | | ν=0.0 | 6.5037 | 12.6329 | Exclude m=0.04÷12 Gev/c$^2$ $10_{-8}^{+25}$ Ellis,93B; $10_{-8}^{+60}$ Novikov, 93B; |
| **ME** | 8.083 | 19.516 | 0.5 | 7.734 | 17.865 | |
| **TR** | 9.191 | 25.228 | 1.0 | 9.197 | 25.265 | $25_{-19}^{+275}$ Ellis, 92E |
| **V** | 11.050 | 36.468 | 1.5 | 10.938 | 35.731 | 35.4 ± 5 Abreu, 92J; $35_{-26}^{+205}$ Ellis, 94 |
| **E** | 12.992 | 50.418 | 2.0 | 13.007 | 50.531 | $50_{-0}^{+353}$ Renton, 92 |
| **(U)** | 15.512 | 71.865 | 2.5 | 15.468 | 71.462 | $73_{-13}^{+178}$ Blondel, 93 |
| **MA** | 16.038 | 76.823 | | | | |
| | 16.5 | 80.918 | | | | $W^\pm$: M=80.84 ± 0.22 ± 0.83 Alitti $N=(2\pi M/M_*)^{1/2}$ = 16.452 $W^\pm$: M=79.91 ± 0.39 Abe N=16.357 |
| | 17.5 | 91.024 | | | | $Z^0$ : M=91.187 ± 0.007 $N=(2\pi M/M_*)^{1/2}$ = 17.473 |
| | 18.5 19.0 | 101.588 107.297 | 3.0 | 18.395 | 101.063 | 103.7 L3 Collaboration [Felcini, 2000] 108.9 L3 Collaboration |
| **(NE)** | 19.431 19.5 | 112.760 [illegible] | | | | 114.5 L3 Collaboration [Felcini, 2000; Tully 2000;Acciarry et al., 2000] |

## Table 2

### MASS SPECTRUM: STATIONARY STATES - $G^{[3]}$ Shell

| T H E O R Y | | | | E X P E R I M E N T |
|---|---|---|---|---|
| | | General Dichotomy | | |
| **States** | **Mass** $M=M_*(N^2/2\pi)$ | **States** $\nu$ | **Mass** $M=M_*\times(N_\nu^2/2\pi)$, $N_\nu= N_{\nu=0}\, 2^{\nu/2}$, $N_{\nu=0} = 6.5$ | **MASS M** |
| | [Tev/c²] | | [Tev/c²] | [Tev/c²] |
| **TR***  | $M_*$=0.339 | | $M_*$=0.339 | **L H C** |
| | | **0.0** | 0.16984 | |
| **ME** | 0.26238 | **0.5** | 0.24019 | |
| **TR** | 0.33918 | **1.0** | 0.33968 | |
| **V** | 0.49030 | **1.5** | 0.48038 | |
| **E** | 0.67784 | **2.0** | 0.67936 | |
| **(U)** | 0.96619 | **2.5** | 0.96077 | |
| **MA** | 1.03285 | | | |
| | | **3.0** | 1.35873 | |
| **(NE)** | 1.516 | | | |
| **CE** | 1.87591 | **3.5** | 1.92154 | |
| **(P)** | 1.98512 | | | |
| | | **4.0** | 2.71747 | |

**Table 3**

## MASS SPECTRUM: STATIONARY STATES - $G^{[4]}$ Shell

| THEORY | | | | EXPERIMENTS |
|---|---|---|---|---|
| | | General Dichotomy | | |
| **States** | **Mass** $M=M_*(N^2/2\pi)$ | **States** $\nu$ | **Mass** $M=M_*\times(N_\nu^2/2\pi)$, $N_\nu=N_{\nu=0}\,2^{\nu/2}$, $N_{\nu=0}=6.5$ | **MASS M** |
| | [Tev/c²] | | [Tev/c²] | [Tev/c²] |
| **TR$_*$** | $M_*$=0.339 | | $M_*$=0.339 | **LHC** |
| | | **0.0** | 2.283 | |
| **ME** | 3.527 | **0.5** | 3.229 | |
| **TR** | 4.560 | **1.0** | 4.566 | |
| **V** | 6.591 | **1.5** | 6.458 | |
| **E** | 9.113 | **2.0** | 9.133 | |
| **(U)** | 12989 | **2.5** | 12.917 | |
| **MA** | 13.886 | | | |
| | | **3.0** | 18.267 | |
| **(NE)** | 20.381 | | | |
| **CE** | 25.220 | **3.5** | 25.834 | |
| **(P)** | 26.688 | | | |
| | | **4.0** | 36.534 | |